\documentclass[12pt
]{article}
\usepackage[dvips]{geometry,color,graphicx}
\usepackage{amsmath}
\usepackage{amssymb}
\usepackage{bm}
\usepackage{float}
\usepackage[toc,page]{appendix}

\begin{document}
\title{Coupling between Ion-Acoustic Waves and Neutrino Oscillations}

\author{Fernando Haas and Kellen Alves Pascoal
\\ \strut \\ Instituto de F\'{\i}sica \\ Universidade Federal do Rio Grande do Sul \\ Av. Bento Gon\c{c}alves 9500 \\ 91501-970 Porto Alegre, RS, Brasil
\\ \strut \\
Jos\'e Tito Mendon\c{c}a 
\\ \strut \\
IPFN, Instituto Superior T\'ecnico \\ Universidade de Lisboa \\ 1049-001 Lisboa, Portugal \\ \strut \\
Instituto de F\'isica \\ Universidade de S\~ao Paulo \\ 05508-090 S\~ao Paulo, SP, Brasil
}

\date{}

\maketitle

\newpage 

\begin{abstract}
\noindent
The work investigates the coupling between ion-acoustic waves and neutrino flavor oscillations in a non-relativistic electron-ion plasma under the influence of a mixed neutrino beam. Neutrino oscillations are mediated by the flavor polarization vector dynamics in a material medium. The linear dispersion relation around homogeneous static equilibria is developed. When resonant with the ion-acoustic mode, the neutrino flavor oscillations can transfer energy to the plasma exciting a new fast unstable mode in extreme astrophysical scenarios. The growth rate and the unstable wavelengths are determined in typical type II supernovae parameters. The predictions can be useful for a new indirect probe on neutrino oscillations in nature. 

\end{abstract}


\section{Introduction}

The investigation of the neutrino properties is a most relevant issue, in the contexts of elementary particle physics, cosmology and astrophysics. On this regard, the timeliness of neutrino physics is shown by the 2015's Nobel Prize awarded to T. Kajita and A. B. McDonald due to the experimental verification of neutrino flavor oscillations, which in turn are a result of the existence of a neutrino mass. From data on baryon acoustic oscillations and cosmic microwave background \cite{Ade}, the estimated sum of neutrino masses has a small upper bound of $0.23 \, {\rm eV}$. Moreover, the mechanism for the generation of neutrino masses is presently uncertain. In spite of all this, the confirmation of neutrino masses and flavor oscillations shows the incompleteness of the Standard Model, calling for a new description of nature. 

Intense beams of neutrinos are present in astrophysical plasmas, like in the lepton era of the early universe \cite{Tajima} or in connection with the question of neutrino heating in type II supernovae \cite{Bethe, Shukla}. As discussed in \cite{Silva}, for the conditions in type II supernovae, neutrino beams therein are collimated enough in order to drive fast plasma instabilities. Indeed, far from the neutrinosphere,  the neutrinos travel in the radial direction, so that the angular velocity dispersion becomes small. In this whole context, it would be an interesting endeavor, to check about the coupling between neutrino and plasma oscillations. At the minimal level, at least one would possibly identify a new experimental test on the elusive neutrino flavor oscillations dynamics, this time in extreme plasma environments subject to strong neutrino ``winds". 

The effective resonance between neutrino oscillations (which are low-frequency modes in general; see Section IV for numerical estimates) and plasma waves is more likely to occur for slow plasma modes, such as the ion-sound branch. Therefore, we focus on an electron-ion plasma system, coupled to electron and muon neutrino flavors. The same problem has been formulated in the literature, but without allowing for neutrino oscillations \cite{Monteiro}. The influence of flavor oscillations on the neutrino-plasma interactions was first considered in a previous article \cite{PoP}. Here, we generalize  this work, by taking into account a non-zero neutrino beam coherence and searching for ion-sound wave stability near static equilibria. Contributions from the magnetic field, which were recently considered in the frame of a neutrino-MHD model \cite{mhd}, will be ignored.

The article is organized as follows. In Sec. II, we write the basic fluid model for the non-relativistic electron-ion plasma coupled to a two-flavor neutrino mixture, where neutrino oscillations are mediated by the appropriate polarization vector. In Sec. III, the equilibrium state and the linear dispersion relation for the ion-sound mode are derived. Conditions for the resonance with neutrino oscillations and the corresponding instability are determined. In Sec. IV, the growth rate is evaluated for parameters compatible with the supernova 1987A. The unstable wavelengths and the time-scale of the instability are then obtained, showing the dominant role of neutrino oscillations over the traditional neutrino-plasma instability mechanism in this case. Sec. V is dedicated to the conclusions. Finally, the Appendix reports a more detailed calculation of the dispersion relation, which merits some algebra in view of the many involved variables.

\section{Physical Model}

The system is described by an hydrodynamical model for electrons, ions, electron-neutrinos and muon-neutrinos. Denoting $n_{e,i}$ and ${\bf u}_{e,i}$ as respectively the electron (e) and ion (i) fluid densities and velocity fields, one will have the continuity equations 
\begin{equation}
\frac{\partial n_e}{\partial t} + \nabla \cdot (n_e {\bf u}_e) = 0 \,,  \quad \frac{\partial n_i}{\partial t} + \nabla \cdot (n_i {\bf u}_i) = 0 \,, \label{eq01} 
\end{equation}
together with the (non-relativistic) electron force equation
\begin{equation}
m_e\left(\frac{\partial}{\partial t} + {\bf u}_e\cdot \nabla \right)\,{\bf u}_e = - \kappa_B T_e \,\frac{\nabla n_e}{n_e} + e\,\nabla\phi + \sqrt{2}\,G_F\,({\bf  E}_\nu + {\bf  u}_{e}\times{\bf  B}_\nu) \,, \label{eq02}
\end{equation}
and cold ions force equation 
\begin{equation}
m_i\left(\frac{\partial}{\partial t} + {\bf u}_i\cdot \nabla \right)\,{\bf u}_i = - e\,\nabla\phi \,. \label{eqi02} 
\end{equation}
In Eqs. (\ref{eq02}) and (\ref{eqi02}), $m_{e,i}$ are the electron (charge $-e$) and ion (charge $+e$) masses, $\kappa_B$ is Boltzmann's constant, $T_e$ is the electron fluid temperature (assuming an isothermal equation of state, appropriate to slow dynamics), and $\phi$ is the scalar potential. Moreover, $G_F$ is Fermi's coupling constant, and ${\bf  E}_\nu, {\bf  B}_\nu$ are effective neutrino electric and magnetic fields given by
\begin{equation}
 {\bf E}_\nu = - \nabla N_e - \frac{1}{c^2}\,\frac{\partial}{\partial t}\,(N_e {\bf  v}_e) \,, \quad {\bf B}_\nu = \frac{1}{c^2}\,\nabla  \times (N_e {\bf  v}_e) \,, \label{eq04}
\end{equation}
where $N_e, {\bf  v}_e$ are the electron-neutrino fluid density and velocity field and $c$ the speed of light. In this work we consider electrostatic excitations described by Poisson's equation
\begin{equation}
\label{poi}
\nabla^2\phi = \frac{e}{\varepsilon_0}\,(n_e - n_i) \,,
\end{equation}
where $\varepsilon_0$ the vacuum permittivity constant. Notice that the Fermi weak force couples only to electrons (leptons), while ions (baryons) are not directly influenced by it. In addition, frequently the treatment of ion-acoustic waves assumes inertialess electrons. However, here $m_e$ is keep on the left-hand side of Eq. (\ref{eq02}) for convenience, but eventually we let $m_e/m_i \approx 0$ (details in the Appendix).  

To investigate the coupling between plasma and neutrino oscillations, we shall consider for simplicity two-flavor neutrino oscillations denoting $N_\mu, {\bf  v}_\mu$ as the muon-neutrino fluid density and velocity field. In this context, one has 
\begin{eqnarray}
 \frac{\partial N_e}{\partial t} + \nabla \cdot (N_e {\bf v}_e) &=& \frac{1}{2}\,N\,\Omega_0\, P_2 \,, \label{eq05} \\
 \frac{\partial N_\mu}{\partial t} + \nabla \cdot (N_\mu {\bf v}_\mu) &=& -\,\frac{1}{2}\,N\,\Omega_0\, P_2 \,, \label{eq06}
\end{eqnarray}
where $N = N_e + N_\mu$ is the total neutrino fluid density and $P_2$ pertains to the quantum coherence contribution in a flavor polarization vector ${\bf  P} = (P_1, P_2, P_3)$. Besides, 
$\Omega_0 = \omega_0 \sin 2\theta_0$, where $\omega_0 = \Delta m^2 c^4/(2\,\hbar\,{\cal E}_0)$ with $\Delta m^2$ being the squared neutrino mass difference. In addition, ${\cal E}_0$ is the neutrino spinor's energy in the fundamental state and $\theta_0$ is the neutrino oscillations mixing angle. The right-hand sides on Eqs. (\ref{eq05}) and (\ref{eq06}) show the contribution from neutrino oscillations, to the electron and muon neutrino density rate of change, while the convective terms on the left-hand sides are due to the neutrino flows. Note that the global neutrino population is preserved, since
\begin{equation}
\frac{d}{dt} \int (N_e + N_\mu)\,d{\bf  r} = - \int \nabla\cdot\left(N_e {\bf v}_e + N_\mu {\bf v}_\mu\right)\,d{\bf r}=    0 \,,
\label{cons}
\end{equation}
assuming decaying or periodic boundary conditions for instance. 

Denoting  ${\bf p}_e = \mathcal{E}_e {\bf v}_e/c^2, \,{\bf p}_\mu = \mathcal{E}_\mu {\bf v}_\mu/c^2$ as the electron and muon neutrino relativistic momenta, where $\mathcal{E}_e, \mathcal{E}_\mu$ are the corresponding neutrino beam energies, one will have the neutrino force equations, 
\begin{eqnarray}
 \frac{\partial {\bf p}_e}{\partial t} + {\bf v}_e \cdot \nabla {\bf p}_e &=& \sqrt{2}\,G_F \left(
- \nabla n_e - \frac{1}{c^2}\,\frac{\partial}{\partial t}\,(n_e{\bf u}_e) + \frac{{\bf v}_e}{c^2} \times \left[\nabla\times(n_e{\bf u}_e)\right]
\right) \,, \label{eq07} 
\\
 \frac{\partial {\bf p}_\mu}{\partial t} + {\bf v}_\mu \cdot \nabla {\bf p}_\mu &=& 0 \,. \label{eq08}
\end{eqnarray}
As discussed elsewhere \cite{sil1, Brizard}, neutrino-plasma interactions can be derived from a Lagrangian formalism, at least when flavor oscillations are absent. A similar neutrino-plasma fluid model has been also put forward for Langmuir waves \cite{Serbeto}, but without flavor oscillations. 

Finally, the time-evolution of the flavor polarization vector ${\bf P} = (P_1, P_2, P_3) $ in a material medium is given \cite{Suekane, Raffelt} by 
\begin{equation}
 \frac{\partial P_1}{\partial t} = -\Omega(n_e)P_2 \,, \quad 
 \frac{\partial P_2}{\partial t} = \Omega(n_e)P_1 - \Omega_0 P_3 \,, \quad 
 \frac{\partial P_3}{\partial t} = \Omega_0 P_2 \,, \label{eqx10}
\end{equation}
where $\Omega(n_e)= \omega_0 [\cos 2 \theta_0 - \sqrt{2}\,G_F\, n_e/(\hbar\omega_0)]$. In a given point of space, one has $\partial|{\bf P}|^2/\partial t = 0$. However, the neutrino oscillations characteristics change in space and time due to the fluctuations of the electrons density. 

The proposed description provides the link between two previous theories: 1) the well known neutrino mass oscillations model, which is in particular a successful approach to solve the solar neutrino deficit problem \cite{Suekane, Raffelt}; 2) the neutrino-plasma coupling model describing the neutrino gas evolution in dense plasmas \cite{Bethe, Shukla, Silva}. The link between these models is established in the neutrino continuity equations (\ref{eq05}) and (\ref{eq06}), where the number densities $N_{e,\mu}$ are affected by the oscillations through the coherence $P_2$, as well as by means of the weak field which is affected by $N_e$ and ${\bf v}_e$ in Eq. (\ref{eq04}). Although a more fundamental theory could be conceived, the proposed quantum fluid model gives an efficient alternative to the unified treatment of neutrino mass oscillations and neutrino-plasma interactions, significantly generalizing the previous work \cite{PoP}.

To summarize, the model comprises the quantities $n_{e,i}, {\bf u}_{e,i}$ (the electron and ion fluid densities and velocity fields), $\phi$ (the electrostatic potential), $N_{e,\mu}$ and ${\bf v}_{e,\mu}$ (the electron and muon neutrino fluid densities and velocity fields) and the three components  of the flavor polarization vector. Taking into account all components, these are 20 variables for a system of 20 equations defined in Eqs. (\ref{eq01})-(\ref{eqi02}), (\ref{poi})-(\ref{eq06}) and (\ref{eq07})-(\ref{eqx10}).

\subsection{Fixed Homogeneous Medium}

For the sake of reference, it is useful to briefly recall the properties of neutrino oscillations in a fixed homogeneous medium where in particular $n_e = n_0, {\bf v}_e = {\bf v}_\mu = 0, \nabla = 0$. In this case, from Eqs. (\ref{eq05}) and (\ref{eq06}) one has 
\begin{eqnarray}
 \frac{\partial N_e}{\partial t} = \frac{1}{2}\,N_0\,\Omega_0\, P_2 \,, \quad \frac{\partial N_\mu}{\partial t} = - \, \frac{1}{2}\,N_0\, \Omega_0\, P_2 \,,
\end{eqnarray}
where $N_0 = N_e + N_\mu$ is not only globally but also locally constant. The time-evolution of the  
flavor polarization vector is then given by 
\begin{equation}
 \frac{\partial P_1}{\partial t} = -\Omega(n_0)P_2 \,, \quad 
 \frac{\partial P_2}{\partial t} = \Omega(n_0)P_1 - \Omega_0 P_3 \,, \quad 
 \frac{\partial P_3}{\partial t} = \Omega_0 P_2 \,, \label{eq17}
\end{equation}
where $\Omega(n_0) = \omega_0 [\cos 2 \theta_0 - \sqrt{2}\,G_F\, n_0/(\hbar\omega_0)]$. It is easy to obtain 
\begin{equation}
\dddot{\bf P} + \Omega_\nu^2\,\dot{\bf P} = 0 \,,
\end{equation}
where $\Omega_\nu$ denotes the eigen-frequency of two-flavor neutrino oscillations in this case, given by 
\begin{equation}
\label{omeganeu}
\Omega_\nu^2 = \Omega^2(n_0)+\Omega_0^2 \,.
\end{equation}
Therefore, obviously we have oscillating solutions $\sim \exp(\pm i\Omega_\nu t)$, around the fixed point
\begin{equation}
\label{peq}
P_1 = \frac{\Omega_0}{\Omega_\nu} \,, \quad P_2 = 0 \,, \quad P_3  = \frac{\Omega(n_0)}{\Omega_\nu} = \frac{N_{e0}-N_{\mu 0}}{N_0} \,,
\end{equation}
where $N_{e0}, N_{\mu 0}$ are the electron and muon equilibrium neutrino fluid densities. The last equality in Eq. (\ref{peq}) is due to the identification $P_3 = (N_e - N_\mu)/N$ assumed to hold in the homogeneous case. For simplicity we have set  $|{\bf P}| = 1$ in equilibrium, corresponding to a pure state (in general, $|{\bf P}| < 1$). 

\section{Linear waves}

Differently from \cite{PoP}, the present approach focus the stability around homogeneous, static equilibrium and not around a dynamic, time-dependent equilibrium representing the neutrino oscillations. As we shall verify, the reformulation allows a more precise understanding of the coupling between plasma and neutrino oscillations. 
Back to the general system, at first consider the homogeneous static equilibrium for Eqs. (\ref{eq01})-(\ref{eqi02}), (\ref{poi})-(\ref{eq06}) and (\ref{eq07})-(\ref{eqx10}), given by 
\begin{eqnarray}
n_e &=& n_i = n_0 \,, \quad {\bf u}_e = {\bf u}_i = 0 \,, \quad \phi = 0 \,, \nonumber \\
N_e &=& N_{e0} \,, \quad N_\mu = N_{\mu 0} \,, \quad {\bf v}_e = {\bf v}_{\mu} = {\bf v}_0  \,, \label{homo}
\end{eqnarray}
together with Eq. (\ref{peq}) for the equilibrium flavor polarization vector. After linearization around the equilibrium for plane wave perturbations $\sim \exp[i({\bf k}\cdot{\bf r} - \omega t)]$ and performing a lengthy calculation detailed in the Appendix, we get
\begin{eqnarray}
 (\omega - {\bf k} \cdot {\bf v}_{0}) \,\delta N_e &=& \frac{\sqrt{2}\,G_F\, N_{e0}}{\mathcal{E}_{0}\,(\omega - {\bf k} \cdot {\bf v}_{0})} 
 \left(c^2\,k^2 - ({\bf k} \cdot {\bf v}_{0})^2\right) \delta n_e \nonumber \\ &+& \frac{i}{2} \,N_0\, \Omega_0\,\delta P_2 \label{eqo} \,, 
\end{eqnarray}
and 
\begin{eqnarray}
\Bigl(\omega^2 - c_s^2 k^2 + \frac{2\,G_{F}^2\,N_{e0}\,n_0\,\omega\,\left(c^2 k^2 - ({\bf k}\cdot{\bf v}_0)^2\right)}{m_i\,c^2\,\mathcal{E}_{0}\,(\omega - {\bf k}\cdot{\bf v}_0)}\Bigr)\,\delta n_e \nonumber \\ = \frac{\sqrt{2}\,G_F n_0}{m_i\,c^2}(c^2 k^2 - \omega\,{\bf k}\cdot{\bf v}_0)\,\delta N_e \,, \label{iaw}
\end{eqnarray}
where $c_s = \sqrt{\kappa_B T_e/m_i}$ is the ion-acoustic speed and $\delta n_e, \delta N_e$ are respectively the electron and electron neutrino fluid densities perturbations, while $\delta P_2$ is the perturbation of the equilibrium quantum coherence. Equation (\ref{eqo}) shows the effect of neutrino oscillations by means of the $\sim \delta P_2$ term. For simplicity, we have assumed $\omega$ much smaller than the ion plasma frequency $\omega_{pi} = \sqrt{n_0\,e^2/(m_i\,\varepsilon_0)}$. In addition, in the present setting, $\mathcal{E}_0 = \mathcal{E}_{e0} \approx \mathcal{E}_{\mu 0}$ is the equilibrium quasi-mono-energetic neutrino beam energy.

It is easy to obtain from the system (\ref{eqx10}) the result 
\begin{equation}
\label{eqpx2} 
\delta\,P_2 = - i \,\frac{\sqrt{2}\,\Omega_0\,\omega\,G_F\,\delta n_e}{(\omega^2 - \Omega_\nu^2)\,\hbar\,\Omega_\nu} \,.
\end{equation}
Clearly the neutrino oscillations term is more relevant for low-frequency waves such that $\omega \approx \Omega_\nu$, as expected on physical grounds. 

Inserting Eq. (\ref{eqpx2}) into Eqs. (\ref{eqo}) and (\ref{iaw}) one then find the dispersion relation 
\begin{eqnarray}
\omega^2 = c_s^2\,k^2 + \frac{\Delta_e\,c^2\,k^2\,\Lambda(\theta)\,(c^2\,k^2 - \omega^2)}{(\omega-{\bf k}\cdot{\bf v}_0)^2} + \frac{\Delta\,\Omega_0^2\,\omega\,\mathcal{E}_0\,(c^2\,k^2 - \omega\,{\bf k}\cdot{\bf v}_0)}{2\,\hbar\,\Omega_\nu\,(\omega - {\bf k}\cdot{\bf v}_0)\,(\omega^2 - \Omega_{\nu}^2)} \,, \label{disp}
\end{eqnarray}
where 
\begin{equation}
\Delta_e = \frac{2\,G_F^2\,N_{e0}\,n_0}{m_i\,c^2\,\mathcal{E}_0} \,, \quad \Delta = \frac{2\,G_F^2\,N_0\,n_0}{m_i\,c^2\,\mathcal{E}_0} \,, \quad \Lambda(\theta) = \left(1 - \frac{v_0^2}{c^2}\right)\,\cos^{2}\theta + \sin^{2}\theta \,,
\end{equation}
with ${\bf k}\cdot{\bf v}_0 = k\,v_0\,\cos\theta$. 

The last term in the right-hand side of Eq. (\ref{disp}) is due to neutrino oscillations. Without this contribution and with $N_{e0} = N_0$, one would regain Eq. (13) of \cite{Monteiro}, taking into account $c_s \ll c$, which is necessary since $c_s \ll v_T = \sqrt{\kappa_B\,T_e/m_e} \ll c$ for non-relativistic electrons. In addition, for simplicity it is assumed that $\omega \ll \omega_{pi}$, to focus on the ion-acoustic rather than on the ionic branch of the dispersion relation. 

\section{Instability of ion-acoustic waves driven by neutrino oscillations}

From Eq. (\ref{disp}) is is apparent that typically the neutrino contribution (with or without neutrino oscillations) is a perturbation to the ion-acoustic waves, due to the small value $G_F = 1.45 \times 10^{-62} \, {\rm J.\,m^3}$ of the Fermi constant. Moreover, the neutrino effect without taking into account neutrino oscillations has been carried out in \cite{Monteiro}. Hence, we focus on the case
\begin{equation}
\label{drc}
\omega \approx c_s\,k = \Omega_\nu = {\bf k}\cdot{\bf v}_0 \,,
\end{equation}
which improves the last term in the right-hand side of the dispersion relation (\ref{disp}), arising from the flavor oscillations. In passing, we note that a separate analysis shows the need of the additional, beam-resonance condition $\omega \approx {\bf k}\cdot{\bf v}_0$ to produce significant corrections to the usual ion-acoustic wave. 

Therefore, we assume Eq. (\ref{drc}) and set 
\begin{equation}
\omega = \Omega_\nu + \delta\omega \,, \quad |\delta\omega| \ll \Omega_{\nu} \,,
\end{equation}
in Eq. (\ref{disp}), to find for ultra-relativistic neutrinos ($v_0 \approx c$) the result 
\begin{equation}
(\delta\omega)^3 = \frac{\Delta_e}{2}\,\left(\frac{c}{c_s}\right)^4\Omega_{\nu}^3 + \frac{G_{F}^2\,N_0\,n_0\,\Omega_{0}^2}{4\,\hbar\,\kappa_B\,T_e}   \,. \label{modd}
\end{equation}
The unstable mode corresponds to a growth rate $\gamma = {\rm Im}(\delta\omega) > 0$ given by 
\begin{equation}
\gamma = \frac{\sqrt{3}}{2}\,\left(\frac{\Delta_e}{2}\,\left(\frac{c}{c_s}\right)^4\Omega_{\nu}^3 + \frac{G_{F}^2\,N_0\,n_0\,\Omega_{0}^2}{4\,\hbar\,\kappa_B\,T_e}\right)^{1/3} \,. \label{gam}
\end{equation}
The neutrino oscillations effect is now contained in the last term inside the cubic root in Eq. (\ref{modd}).

It is convenient to define 
\begin{equation}
\gamma_\nu = \frac{\sqrt{3}}{2}\,\left(\frac{\Delta_e}{2}\,\left(\frac{c}{c_s}\right)^4\right)^{1/3}\Omega_{\nu} \,, \quad \gamma_{\rm osc} = \frac{\sqrt{3}}{2}\,\left(\frac{G_{F}^2\,N_0\,n_0\,\Omega_{0}^2}{4\,\hbar\,\kappa_B\,T_e}\right)^{1/3} \,, \label{gs}
\end{equation}
so that from Eq. (\ref{gam}) one has $\gamma^3 = \gamma_{\nu}^3 + \gamma_{\rm osc}^3$. The quantity $\gamma_\nu$ corresponds to the usual neutrino-plasma coupling, without accounting for the neutrino oscillations. The flavor conversion effect is associated to $\gamma_{\rm osc}$. 

We have a convenient setting for the evaluation of the neutrino oscillations effect. We get
\begin{equation}
\left(\frac{\gamma_\nu}{\gamma_{\rm osc}}\right)^3 = \frac{4\,\hbar}{\mathcal{E}_0}\,\frac{\Omega_\nu^3}{\Omega_0^2}\,\left(\frac{c}{c_s}\right)^2\frac{N_{e0}}{N_0} \,. \label{comp}
\end{equation}
It is interesting to note that for a muonic-neutrino beam ($N_{e0} = 0 \,, N_{\mu 0} = N_0$) the neutrino oscillations term completely dominates the instability. This is to be expected, since the muon-neutrinos do not couple to the electrons, so that the plasma is affected by the neutrino beams only because of the flavor conversion when some muon-neutrinos gradually become electron-neutrinos in this case.

For the dense plasmas under consideration, it can be safely assumed that $\Omega(n_0) \approx - \sqrt{2}\,G_F\,n_0/\hbar$, a negative value corresponding to an inverted neutrino mass hierarchy. Moreover, $\Omega_\nu \approx |\Omega(n_0)|$, so that Eq. (\ref{comp}) can be accurately replaced by 
\begin{equation}
\left(\frac{\gamma_\nu}{\gamma_{\rm osc}}\right)^3 \approx \frac{8\,\sqrt{2}\,(G_F \, n_0)^3}{\mathcal{E}_0 \,\left(\hbar\,\omega_0\sin(2\theta_0)\right)^2}\,\left(\frac{c}{c_s}\right)^2\frac{N_{e0}}{N_0} = \frac{35.16}{\kappa_B\,T_e}\,\frac{N_{e0}}{N_0} \,,
\end{equation}
where the last equality assumes $\mathcal{E}_0 = 10\,{\rm MeV}, n_0 = 10^{35}\,{\rm m}^{-3}$ and where $\kappa_B T_e$ is measured in ${\rm keV}$. Therefore, if $\kappa_B\,T_e > 35.16\,{\rm keV}$, the neutrino-oscillations-driven instability will certainly dominates the usual neutrino-plasma coupling instability mechanism, as long as the ion-acoustic wave matches the neutrino oscillations, or $\omega \approx \Omega_\nu$. 

\begin{figure}
\centering
\includegraphics[width=300pt
]{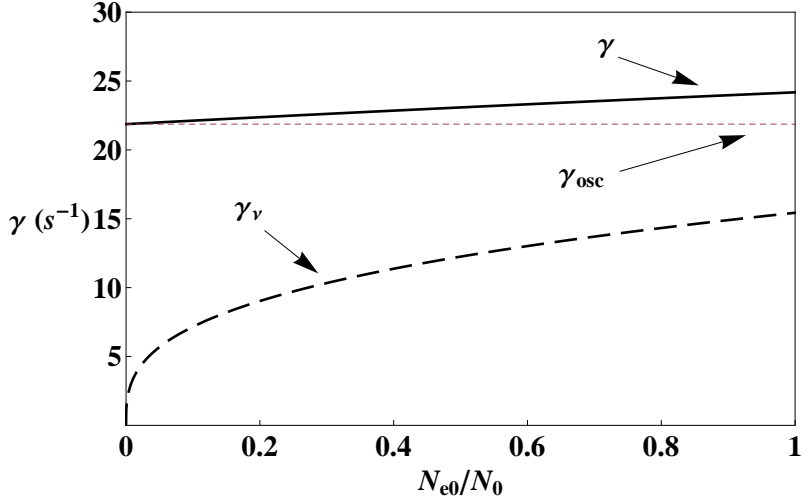}
\caption{Continuous line: growth rate $\gamma$ from Eq. (\ref{gam}) in terms of the normalized electron-neutrino population. Line-dashed curve: the growth rate $\gamma_\nu$ which would take place without neutrino oscillations. Horizontal dot-dashed line: the growth rate 
$\gamma_{\rm osc}$ due uniquely to neutrino oscillations; see Eq. (\ref{gs}). Parameters: ${\cal E}_0 = 10 \,{\rm MeV}, N_0 = 10^{41}\,{\rm m}^{-3}, n_0 = 10^{35} \, {\rm m}^{-3}$, $\kappa_B T_e = 100 \,{\rm keV}$.}
\label{figure}
\end{figure}

We evaluate the growth rate in type II core-collapse supernovae scenarios, as for the supernova SN1987A with a neutrino burst of $10^{58}$ neutrinos of all flavors and energy between $10-15$ MeV \cite{Hirata}. First consider $\Delta m^2\,c^4 = 3 \times 10^{-5} \,({\rm eV})^2 \,,\sin(2\theta_0) = 10^{-1}$, which are suitable parameters to solve the solar neutrino problem \cite{Raffelt}. Also take ${\cal E}_0 = 10 \,{\rm MeV}, \kappa_B T_e = 100 \,{\rm keV}, N_0 = 10^{41}\,{\rm m}^{-3}, n_0 = 10^{35} \, {\rm m}^{-3}$ For the described parameters, we have $\omega \approx \Omega_\nu = 1.94 \times 10^7 \,{\rm rad}/s$, much smaller than $\omega_{pi} = 4.16 \times 10^{17}\,{\rm rad}/s$ as requested. Moreover, $c_s = 3.10 \times 10^6 \, {\rm m/s} \ll c$ and $k = \Omega_{\nu}/c_s = 6.28 \,{\rm m}^{-1}$, corresponding to a wavelength $\lambda = 2\pi/k = 1.00 \, {\rm m}$. Finally, $\gamma_{\rm osc} = 21.87 \, {\rm s}^{-1}$ and the maximal growth rate is $\gamma_{\rm max} = 24.18 \, {\rm s}^{-1}$. Therefore,  $1/\gamma_{\rm max} \sim 0.04 \, {\rm s}$, which is fast enough to drive a supernova explosion whose accepted characteristic time is around 1 second, as well as of the same order of magnitude of other neutrino-plasma instability growth rates (see \cite {Silva} for a review, and more recent results in \cite{mhd}).   Figure \ref{figure} shows the numerical value of instability growth rate, for different normalized electron-neutrino populations. One sees that the neutrino-oscillations-driven instability mechanism is always dominant over the usual neutrino-plasma coupling in this case. In comparison to previous studies on instabilities due to neutrino-plasma interactions \cite{Bethe}-\cite{Serbeto}, the results show a much longer wavelength, which can be presumed to be a welcome feature for practical observations. All in all, the estimates provide an indirect signature of flavor conversion in terms of the destabilization of the ion-acoustic waves resonant with the neutrino oscillations.

In passing we note that we have a filamentation-like instability with an almost orthogonal propagation since $\cos\theta = c_s/c \ll 1$. Moreover Landau damping is not an issue for ion-acoustic waves as long as $T_i \ll T_e$ where $T_i$ is the ions fluid temperature. In addition, the parameters of Figure \ref{figure} are still non-relativistic to a good approximation, with a thermal relativistic factor $\gamma_T = 1/\sqrt{1- \kappa_B\,T_e/(m_e\,c^2)} = 1.12 \approx 1$. Finally, the plasma can be also taken as non-degenerate to reasonable accuracy, with a Fermi energy $E_F = \hbar^2 (3\,\pi^2\,n_0)^{2/3}/(2\,m_e) = 78.76 \, {\rm keV} < \kappa_B \, T_e$. To include degeneracy effects, the equation of state of an isothermal degenerate Fermi gas would be required, with the net effect \cite{Haas1, Haas2} of the replacement of $c_s$ by a generalized ion-acoustic velocity $C_s$ given by
\begin{equation}
C_s = \left(\frac{\kappa_B\,T_e}{m_i}\,\frac{L_{3/2}[-\exp(\beta\mu_0)]}{L_{1/2}[-\exp(\beta\mu_0)]}\right)^{1/2} \,,
\end{equation}
where $\beta = 1/(\kappa_B\,T_e)$,  $L_{j}(- z)$ is the polylogarithm function \cite{Lewin} of index $j$ with $z = \exp(\beta\mu_0)$, and $\mu_0$ is the equilibrium chemical potential. 
When $j > 0$, the polylogarithm function 
can be defined as 
\begin{equation}
{\rm L}_{j }(- z )=-\frac{1}{\Gamma (j )}\int_{0}^{\infty }
\frac{s^{j -1}}{1 + e^{s}/z}ds \,, 
\end{equation}
where $\Gamma(j )$ is the gamma function. 
The equilibrium chemical potential $\mu_0$ is related to the
equilibrium density $n_0$ through 
\begin{equation}  
- \,\frac{n_0}{{\rm L}_{3/2}(-e^{\beta \mu_0 })}\left(\frac{\beta m_{e}}{%
2\pi} \right) ^{3/2} = 2\left( \frac{m_{e}}{2\pi \hbar }\right) ^{3}  \,.
\end{equation}
For the chosen parameters, we find $\beta\mu_0 = - 0.58$ and $C_s = 1.09 \,c_s$, so that the degeneracy does not affect too much the results. However, obviously other specific cases should be checked with care, both for relativistic and Fermi pressure effects. 

Furthermore, one might wonder about the relevance of quantum diffraction effects, contained in the Bohm potential \cite{book}. This term induces \cite{Garcia} a correction of the order $\sim \hbar^2 k^4/m_{e}^2$ to $\omega^2 \approx c_{s}^2\,k^2$, so that we can estimate a dimensionless quantum diffraction parameter defined by
\begin{equation}
\left(\frac{\hbar\,k}{m_{e}\,c_{s}}\right)^2 = \left(\frac{\hbar\,\Omega_\nu}{m_{e}\,c_{s}^2}\right)^2 \approx 10^{-20} \,.
\end{equation}
Therefore quantum diffraction effects are completely negligible at least for the long, macroscopic wavelengths under consideration.

Finally, it should be realized that an essential ingredient of the treatment is the anisotropic part of the neutrino velocities distribution corresponding to the neutrino beam. The radial flux from the exterior of the neutrinosphere, at long distances from the center, has a small angular spread so that it can be treated as a collimated beam \cite{Silva}. Mechanisms for neutrino velocities anisotropy have been discussed elsewhere \cite{Laming}.

\section{Conclusion}

In this work we reformulate and generalize the treatment of \cite{PoP} in several ways, namely, (a) allowing more general fluctuations of the neutrino fluid densities $N_{e,\mu}$, so that the total neutrino fluid density $N = N_e + N_\mu$ can locally fluctuate; 
(b) performing the linear stability analysis around static, homogeneous equilibrium solutions for the plasma plus mixed neutrinos system, instead of considering dynamic equilibria. (c) The reformulation allows to treat the coupling between plasma and neutrino oscillations without any restrictions. By comparison, in \cite{PoP} more precise analytic results were available only for the case of vanishing quantum coherence, which is the less interesting situation. In this way there is a very clear significant improvement to the literature, where the dispersion relation (\ref{disp}) generalize the results in \cite{Monteiro} by taking into account the impact of neutrino oscillations on ion-acoustic plasma waves. 

The present findings can be helpful for independent experimental verifications of the neutrino mass, in connection with the destabilization of ion-acoustic waves coupled to flavor oscillations in extreme astrophysical settings. The precise wavelengths and linear growth rate for the new instability mechanism have been identified. The neutrino oscillations free energy source has been found to be the generic dominant influence in such situations, in comparison to the traditional neutrino-plasma interaction. 

\vspace{.5cm}

{\bf  Acknowledgments}: 

F.~H.~ and J.~T.~M.~ acknowledge the support by Con\-se\-lho Na\-cio\-nal de De\-sen\-vol\-vi\-men\-to Cien\-t\'{\i}\-fi\-co e Tec\-no\-l\'o\-gi\-co (CNPq) and EU-FP7 IRSES Programme (grant 612506 QUANTUM PLASMAS FP7-PEOPLE-2013-IRSES), and K.~A.~P.~ack\-now\-ledges the support by Coordena\c{c}\~ao de Aperfei\c{c}oamento de Pessoal de N\'{\i}vel Superior (CAPES). 

\appendix

\section{Derivation of Eqs. (\ref{eqo}) and (\ref{iaw})}

We linearize the model equations (\ref{eq01})-(\ref{eqi02}), (\ref{poi})-(\ref{eq06}) and (\ref{eq07})-(\ref{eqx10}) around the equilibrium (\ref{homo}), for plane wave-perturbations $\sim \exp[i({\bf k}\cdot{\bf r} - \omega\,t)]$, denoting perturbed quantities with a $\delta$ in front of it. So for instance $n_e = n_0 + \delta n_e \,\exp[i({\bf k}\cdot{\bf r} - \omega\,t)]$. Performing straightforward operations, we find from the electron momentum equation that 
\begin{equation}
\frac{n_0\,m_e\,\omega\,\delta{\bf u}_e}{m_i} = - \,\left(\frac{\omega^2}{k^2} - c_s^2\right)\,{\bf k}\,\delta n_e + \frac{\sqrt{2}\,G_F\,n_0}{m_i\,c^2}\,\left((c^2\,{\bf k} - 
\omega\,{\bf v}_0)\,\delta N_e - \omega\,N_{e0}\,\delta{\bf v}_e\right) \,, \label{me}
\end{equation}
taking into account $\omega \ll \omega_{pi}$, while the electron neutrino momentum equation gives 
\begin{eqnarray}
c^2\,(\omega &-& {\bf k}\cdot{\bf v}_0)\,\delta{\bf p}_e = \mathcal{E}_{e0}\,(\omega - {\bf k}\cdot{\bf v}_0)\,\left[\delta{\bf v}_e + 
\left(1 - \frac{v_0^2}{c^2}\right)^{-1}\frac{{\bf v}_0\cdot\delta{\bf v}_e}{c^2}\,{\bf v}_0\right] \nonumber \\ 
&=&
\sqrt{2}\,G_F\,n_0\left(c^2\,{\bf k}\,\frac{\delta n_e}{n_0} - \omega\delta{\bf u}_e - \,{\bf v}_0\times({\bf k}\times\delta{\bf u}_e)\right) \,. \label{kkk} 
\end{eqnarray}
The first equality in Eq. (\ref{kkk}) comes from ${\bf p}_e = \mathcal{E}_e{\bf v}_{e}/c^2$ and $\mathcal{E}_e = m_{\nu}\,c^2 (1- v_e^2/c^2)^{-1/2}$ assuming a neutrino mass $m_{\nu}$ just for the sake of the calculation (at the end it does not appear). 

Equation (\ref{kkk}) can be solved for $\delta{\bf v}_e$ as
\begin{eqnarray}
\delta{\bf v}_e = \frac{\sqrt{2}\,G_F}{\mathcal{E}_0\,(\omega - {\bf k}\cdot{\bf v}_0)}\,\Bigl[c^2\,{\bf k}\,\delta n_e &-& n_0\,\omega\,\delta{\bf u}_e - \left({\bf k}\cdot{\bf v}_0\,\delta n_e - \frac{n_0\,\omega}{c^2}\,{\bf v}_0\cdot\delta{\bf u}_e\right)\,{\bf v}_0 \nonumber \\
&-& n_0\,\left({\bf v}_0\cdot\delta{\bf u}_e\,{\bf k} - {\bf k}\cdot{\bf v}_0\,\delta{\bf u}_e\right)\Bigr] \,, \label{vv}
\end{eqnarray}
where in the right-hand side it was approximated $\mathcal{E}_{e0} \approx \mathcal{E}_0$. 

Taking the scalar product with ${\bf k}$ and using Eq. (\ref{vv}), the electron momentum equation (\ref{me}) gives
\begin{eqnarray}
\frac{m_e\,\omega^2\,\delta n_e}{m_i} &=& - (\omega^2 - c_s^2 k^2)\,\delta n_e + \frac{\sqrt{2}\,G_F\,n_0}{m_i\,c^2}\,(c^2 k^2 - 
\omega\,{\bf k}\cdot{\bf v}_0)\,\delta N_e \nonumber \\
&-& \frac{2\,G_{F}^2 N_{e0}\,n_0\,\omega}{m_i\,c^2\mathcal{E}_{0}\,(\omega - {\bf k}\cdot{\bf v}_{0})} \Bigl[\Bigl(c^2 k^2 - \omega^2 - ({\bf k}\cdot{\bf v}_{0})^2 \Bigr) \,\delta n_e  \label{q}  \\ &+& \frac{n_0\,\omega}{c^2}\,({\bf k}\cdot{\bf v}_0)\,({\bf v}_{0}\cdot\delta{\bf u}_e) - n_0\,k^2\,{\bf v}_0\cdot\delta{\bf u}_e + n_0\,({\bf k}\cdot{\bf v}_0)\,({\bf k}\cdot\delta{\bf u}_e)\Bigr] \,. \nonumber 
\end{eqnarray}

To proceed, we approximate $\delta{\bf u}_e$ from Eq. (\ref{me}) as
\begin{equation}
\delta{\bf u}_e \approx - \,\frac{m_i}{n_0\,m_e\,\omega}\,\left(\omega^2 - c_s^2 k^2\right)\,\frac{{\bf k}\,\delta n_e}{k^2} \approx \frac{\omega\,{\bf k}\,\delta n_e}{n_0\,k^2} \,, \label{simp}
\end{equation}
to substitute into Eq. (\ref{q}), since the terms containing ${\bf u}_e$ in this equation are already of order 
$\mathcal{O}(G_F^2)$. The last approximation in Eq. (\ref{simp}) follows from the formal classical limit ($G_F = 0$) and is consistent with the linearized electron fluid continuity equation $\omega\,\delta n_e = n_0\,{\bf k}\cdot\delta{\bf u}_e$. 

Inserting Eq. (\ref{simp}) into Eq. (\ref{q}) yields 
\begin{eqnarray}
\frac{m_e\,\omega^2\,\delta n_e}{m_i} &=& - (\omega^2 - c_s^2 k^2)\,\delta n_e + \frac{\sqrt{2}\,G_F\,n_0}{m_i\,c^2}\,(c^2 k^2 - 
\omega\,{\bf k}\cdot{\bf v}_0)\,\delta N_e \\ 
 &-& \frac{2\,G_{F}^2 N_{e0}\,n_0\,\omega}{m_i\,c^2\mathcal{E}_{0}\,(\omega - {\bf k}\cdot{\bf v}_{0})}\Bigl(1 - \frac{\omega^{2}}{c^2\,k^2}\Bigr)\,\Bigl(c^2 k^2 - ({\bf k}\cdot{\bf v}_{0})^2 \Bigr) \,\delta n_e \,, \nonumber 
\end{eqnarray}
which is Eq. (\ref{iaw}) taking into account $\omega^2/(c^2\,k^2) \approx c_{s}^2/c^2 \ll 1$ and $m_e/m_i \ll 1$.

On the other hand, from the electron neutrino continuity equation, we have 
\begin{equation}
(\omega - {\bf k}\cdot{\bf v}_0)\,\delta N_e - N_{e0}\,{\bf k}\cdot\delta{\bf v}_e = \frac{i\,N_0\,\Omega_0\,\delta P_2}{2} \,. \label{xx}
\end{equation}
From Eqs. (\ref{vv}) and (\ref{simp}), we have
\begin{eqnarray}
\delta{\bf v}_e &=& \frac{\sqrt{2}\,G_F}{\mathcal{E}_0\,(\omega -  {\bf k}\cdot{\bf v}_0)}\,\Bigl(1 - \frac{\omega^{2}}{c^2\,k^2}\Bigr)\,\Bigl(c^2 {\bf k} - {\bf k}\cdot{\bf v}_{0} \,\,{\bf v}_0\Bigr)\,\delta n_e \nonumber \\ &\approx& \frac{\sqrt{2}\,G_F}{\mathcal{E}_0\,(\omega -  {\bf k}\cdot{\bf v}_0)}\,\Bigl(c^2 {\bf k} - {\bf k}\cdot{\bf v}_{0} \,\,{\bf v}_0\Bigr)\,\delta n_e \,.
\end{eqnarray}
Inserting the last result into Eq. (\ref{xx}), we eventually find Eq. (\ref{eqo}).

\end{document}